\documentclass[aps,pra,nofootinbib,noindent,superscriptaddress,twocolumn,
]{revtex4}
\usepackage[pdftex]{graphicx}
\usepackage{mathrsfs}
\usepackage[colorlinks]{hyperref}
\usepackage{amsmath}
\usepackage{amsfonts}
\usepackage{amssymb}
\providecommand{\U}[1]{\protect\rule{.1in}{.1in}}

\newtheorem{theorem}{Theorem}
\newtheorem{problem}[theorem]{Problem}
\newtheorem{remark}[theorem]{Remark}

\newcommand{\cc}[1]{\textsc{#1}}

\newcommand{\R}[0]{\mathcal{R}}

\begin{document}
\title{Operational meaning of quantum measures of recovery}

\author{Tom Cooney}
\affiliation{Department of Mathematics, State University of New York at Geneseo, Geneseo, New York 14454, USA}
\author{Christoph Hirche} 
\affiliation{F\'{\i}sica Te\`{o}rica: Informaci\'{o} i Fen\`{o}mens 
Qu\`{a}ntics, Departament de F\'{i}sica, Universitat Aut\`{o}noma de Barcelona, ES-08193 
Bellaterra (Barcelona), Spain}
\author{Ciara Morgan}
\affiliation{School of Mathematics and Statistics,
University College Dublin,
Belfield, Dublin 4, Ireland}
\author{Jonathan~P.~Olson}
\affiliation{Hearne Institute for Theoretical Physics,  Department of Physics and Astronomy, Louisiana State University, Baton Rouge, Louisiana 70803, USA}
\author{Kaushik~P.~Seshadreesan}
\affiliation{Max  Planck  Institute  for  the  Science  of  Light,
Guenther-Scharowsky-Str.~1/  building  24,  91058  Erlangen,  Germany}
\author{John Watrous}
\affiliation{Institute for Quantum Computing and School of Computer Science,
University of Waterloo, Canada}
\affiliation{Canadian Institute for Advanced Research,
Toronto, Canada}
\author{Mark~M.~Wilde}
\affiliation{Hearne Institute for Theoretical Physics,  Department of Physics and Astronomy, Louisiana State University, Baton Rouge, Louisiana 70803, USA}
\affiliation{Center for Computation and Technology,  Louisiana State University, Baton Rouge, Louisiana 70803, USA}
\date{\today}

\begin{abstract}
Several information measures have recently been defined which capture the notion of ``recoverability.''  In particular, the fidelity of recovery quantifies how well one can recover a system $A$ of a tripartite quantum state, defined on systems $ABC$, by acting on system $C$ alone.  The relative entropy of recovery is an associated measure in which the fidelity is replaced by relative entropy.
In this paper, we provide concrete operational interpretations of the aforementioned recovery measures  in terms of a computational decision problem and a hypothesis testing scenario. Specifically, we show that the fidelity of recovery is equal to the maximum probability with which a computationally unbounded quantum prover can convince a computationally bounded quantum verifier that a given quantum state is recoverable. The quantum interactive proof system giving this operational meaning requires four messages exchanged between the prover and verifier, but by forcing the prover to perform his actions in superposition, we construct a different proof system that requires only two messages. The result is that the associated decision problem is in \cc{QIP}(2) and another argument establishes it as hard for \cc{QSZK}
(both classes contain problems believed to be difficult to solve for a quantum computer). We finally prove that the regularized relative entropy of recovery is equal to the optimal Type II error exponent when trying to distinguish many copies of a tripartite state from a recovered version of this state, such that the Type I error is constrained to be no larger than a  constant.
\end{abstract}

\maketitle

\section{Introduction}
There are many facets of quantum science in which the notion of quantum state recovery is deeply embedded.  This is particularly true for quantum error correction \cite{Gai08,LB13} and quantum key distribution \cite{SBCDLP09}, where the primary goal is fundamentally that of recovery.  In the former, the task is to reconstruct a quantum state where some part of the state has undergone noise or loss; in the latter, the task is to keep a message secure against an eavesdropper attempting a similar reconstruction.  In either case, the success or failure of a protocol often hinges on whether a particular state in question is recoverable at all, or if the state is beyond repair. 

A particularly important class of states are those that constitute a Markov chain.  A classical Markov chain can be understood as a memoryless random process, i.e., a process in which the state transition probability depends only on the current state, and not on past states.  If random variables $X$, $Y$, and $Z$ form a classical Markov chain as $X\rightarrow Y \rightarrow Z$, then the classical conditional mutual information $I(X;Z|Y)=0$, where
\begin{equation}
I(X;Z|Y) \equiv H(XY) + H(ZY) - H(Y) - H(XYZ)
\end{equation}
and  $H(X)$ is equal to the Shannon entropy of $X$.  Classical Markov chains model an impressive number of natural processes in physics and many other sciences \cite{Norris97}.

An attempt at understanding a quantum generalization of these ideas was put
forward in \cite{HJPW04}, but it was later realized that these notions 
made sense only in the exact case \cite{ILW08}. That is, in analogy with the classical case mentioned above, the authors of
\cite{HJPW04}\ defined a quantum Markov chain to be a tripartite state
$\rho_{ABC}$\ for which the conditional quantum mutual information (CQMI)
$I(A;B|C)_{\rho}$\ is equal to zero, where
\begin{multline}
I(A;B|C)_{\rho} \equiv H(AC)_\rho + H(BC)_\rho 
-H(C)_\rho\\ - H(ABC)_\rho
\end{multline}
and $H(AC)_\rho$ is equal to the von Neumann entropy of the reduced state $\rho_{AC}$
(and likewise for $H(BC)_\rho$, $H(C)_\rho$, and $H(ABC)_\rho$). However,
the later work in \cite{ILW08} (see also \cite{E15}) demonstrated that large  perturbations of a quantum Markov state as defined in \cite{HJPW04} can sometimes lead only to small increases of the CQMI, calling into question the definition of quantum Markov chains from \cite{HJPW04}.

Meanwhile, it has been known for some time that an equivalent description for the exact case $I(A;B|C)_{\rho} = 0$ exists in terms of recoverability.  The work of Petz
\cite{Petz1986,Petz1988}\ implies that there exists a recovery channel
$\mathcal{R}_{C\rightarrow AC}$ such that $\rho_{ABC}=\mathcal{R}%
_{C\rightarrow AC}(\rho_{BC})$ if and only if $I(A;B|C)_{\rho} = 0$. This is in perfect analogy with the exact
classical case mentioned above:\ for a state satisfying $I(A;B|C)_{\rho} = 0$, one could lose the $A$ system and recover it back
from $C$ alone. In this sense, all correlations between systems $A$ and $B$ are
mediated through system $C$ for quantum Markov chain states. Recoverability in this sense is thus intimately connected to Markovianity and represents a method for handling the approximate case, different from that given in \cite{HJPW04}.

To measure non-Markovianity in the approximate case, the general approach outlined in \cite{SW14} was to quantify the ``distance'' from $\rho_{ABC}$ to its closest recovered version. The main measure on which \cite{SW14} focused was the \textit{fidelity of recovery}, defined as
\begin{equation}
F(A;B|C)_{\rho}\equiv\sup_{\mathcal{R}_{C\rightarrow AC}}F(\rho_{ABC},\mathcal{R}_{C\rightarrow AC}(\rho_{BC})), \label{eq:fidrec}
\end{equation}
where the quantum fidelity is defined as
\begin{equation}
F(\omega,\tau) \equiv \Vert \sqrt{\omega} \sqrt{\tau} \Vert_1^2
\end{equation} for density operators $\omega$ and $\tau$ \cite{U73}. The optimization in \eqref{eq:fidrec} is with respect to quantum channels $\mathcal{R}_{C\rightarrow AC}$\ acting on the $C$ system and producing an output on the $A$ and $C$ systems. A related measure, defined in \cite[Remark 6]{SW14}, is the \textit{relative entropy of recovery}:
\begin{equation}
D(A;B|C)_{\rho} \equiv \inf_{\R_{C\rightarrow AC}}D(\rho_{ABC}\Vert\R_{C\rightarrow AC}(\rho_{BC})).
\end{equation}
The quantum relative entropy is defined as
\begin{equation}
D(\omega\Vert\tau) \equiv \operatorname{Tr}\{ \omega[\log \omega - \log \tau]\}
\end{equation}
 if $\operatorname{supp}(\omega) \subseteq
\operatorname{supp}(\tau)$ and it is equal to $+\infty$ otherwise \cite{U62}.
These are clearly well motivated measures of recovery / non-Markovianity, but hitherto they have been lacking concrete operational interpretations. This is the main question that we address in this paper.

From the main result of \cite{FR14}, which established that
\begin{equation}
I(A;B|C)_{\rho} \geq -\log F(A;B|C)_{\rho},
\end{equation}
it is now understood that the CQMI itself is a measure of non-Markovianity as well. Before \cite{FR14}, an operational interpretation for the CQMI had already been given in \cite{DY08, YD09} as twice the optimal rate of quantum communication needed for a sender to transfer one share of a tripartite state to a receiver (generally shared entanglement is required for this task). Here, the decoder at the receiving end in this protocol plays the role of a recovery channel, an interpretation later used in \cite{BHOS14}. 
 A wave of recent work \cite{BCY11,Winterconj,K13conj,Z14,BSW14,SBW14,LW14,BLW14,DW15,BT15,SOR15,W15,DW15a,Sutter15,Junge15} on this topic has added to and complements \cite{FR14}, solidifying what appears to be the right notion of quantum Markovianity.

It follows from the concerns in recovery applications that one may have to systematically decide whether or not a given tripartite quantum state is recoverable. In this paper, we discuss two concrete scenarios in which this is the case. The first scenario is an experiment involving a single copy of the state $\rho_{ABC}$ and the second involves many copies of such a state---for both settings, the goal is to decide whether a given tripartite state is recoverable.

In more detail, the first scenario asks: given a description of a quantum circuit that prepares a state $\rho_{ABC}$, what is the maximum probability with which someone could be convinced that the state is recoverable? Also, how difficult is the task of deciding if the state meets some criteria of recoverability when $A$ is lost?  We address these questions by defining the associated decision problem, called \cc{FoR} for ``fidelity of recovery.'' Using ideas from quantum complexity theory \cite{W09,VW15}, we show that the fidelity of recovery is equal to the maximum probability with which a verifier can be convinced that $\rho_{ABC}$ is recoverable from $\rho_{BC}$ by acting on system $C$ alone. The quantum interactive proof system establishing this operational meaning for the fidelity of recovery
is depicted in Figure~\ref{fig:for} and
follows intuitively from the duality property of fidelity of recovery, originally established in \cite{SW14}.
It also proves that \cc{FoR} is contained in the complexity class \cc{QIP} \cite{W09,VW15}.

However, the proof system in Figure~\ref{fig:for}
requires the exchange of four messages  between
the verifier and 
the prover, and from a computational complexity theoretic perspective, it is desirable to reduce the number of messages exchanged. In fact, this is certainly possible because a general procedure is known which reduces any quantum interactive proof system to an equivalent one which has only three messages exchanged \cite{QIPAmp00}. In Section \ref{sec:FoR-in-QIP2}, we contribute a different proof system for \cc{FoR} which requires
the exchange of only two messages  between the verifier and the prover. The main idea is that the verifier can force the prover to perform his actions in superposition, and 
the result is that the \cc{FoR} decision problem is in \cc{QIP}(2).
 We also argue that \cc{FoR} is hard for \cc{QSZK} \cite{W02,W09zkqa}, by 
  building on earlier work in \cite{v011a003}.
Note that both \cc{QSZK} and \cc{QIP}(2) contain problems believed to be difficult to solve by a quantum computer.

The second scenario in which we give an operational meaning for a recovery measure is an experiment involving many copies of the state $\rho_{ABC}$. Let $n$ be a large positive integer. Suppose that  either the state $\rho_{ABC}^{\otimes n}$ is prepared or the state $\mathcal{R}_{C^n \to A^n C^n}(\rho_{BC}^{\otimes n})$ is, where $\mathcal{R}_{C^n \to A^n C^n}$ is some arbitrary collective recovery channel acting on all $n$ of the $C$ systems. The goal is then to determine which is the case by performing a collective measurement on all of the systems $A^n B^n C^n$. There are two ways that one could make a mistake in this hypothesis testing setup. The first is known as the Type I error, and it is equal to the probability of concluding that 
$\mathcal{R}_{C^n \to A^n C^n}(\rho_{BC}^{\otimes n})$ was prepared if in fact $\rho_{ABC}^{\otimes n}$ was prepared. The other kind of error is the Type II error. Defining the regularized relative entropy of recovery as follows \cite{BHOS14}:
\begin{equation}
D^{\infty}(A;B|C)_{\rho} \equiv  \lim_{n\rightarrow\infty}\frac{1}{n}D(A^n;B^n|C^n)_{\rho^{\otimes n}},
\end{equation}
we prove that
$D^{\infty}(A;B|C)_{\rho}$ is equal to  
the optimal exponent for the Type II error if the Type I error is constrained to be no larger than a constant $\varepsilon \in (0,1)$. That is, there exists a measurement such that the Type II error goes as  $\approx 2^{-n D^{\infty}(A;B|C)_{\rho}}$ with the Type I error no larger than $\varepsilon$. However, if one tries to make the Type II error decay faster than   $\approx 2^{-n D^{\infty}(A;B|C)_{\rho}}$, then it is impossible to meet the Type I constraint for any  $\varepsilon \in (0,1)$. Thus, our result establishes a concrete operational interpretation of the regularized relative entropy of recovery in this hypothesis testing experiment.
It was previously shown in \cite{BHOS14} that
\begin{equation}
I(A;B|C)_\rho\geq D^\infty(A;B|C)_\rho \geq -\log F(A;B|C)_{\rho},
\end{equation}
but no operational interpretation of $D^\infty(A;B|C)_\rho$ was given there.
In the rest of the paper, we provide more details of these operational interpretations in order to justify them.

\section{Operational meaning of fidelity of recovery}

We now provide an operational interpretation for the fidelity of recovery, by considering the following computational task:
 \begin{problem}[\cc{FoR}] 
 \label{prob:FoR}
Given is a description of a quantum circuit that prepares a tripartite state $\rho_{ABC}$, along with real numbers $\alpha,\beta\in (0,1)$ satisfying $\alpha - \beta \geq [\operatorname{poly}(n)]^{-1}$, for $n$ denoting the circuit size. Promised that either
\begin{align*}
\textrm{YES}:& \  F(A;B|C)_{\rho}\geq\alpha \qquad \textrm{or,}  \nonumber \\
\textrm{NO}: &\  F(A;B|C)_{\rho}\leq\beta,  \nonumber 
\end{align*}
decide which of the above is the case.
\end{problem}
\begin{remark}
The additional assumption that $\alpha-\beta\geq[\operatorname{poly}(n)]^{-1}$ is a common assumption which allows for  amplifying the probability of deciding correctly, by employing error reduction procedures \cite{QIPAmp00,JUW09}.  This kind of assumption is required for most applications in quantum complexity theory \cite{KSV02,VW15}.
\end{remark}

  The computational problem \cc{FoR} is defined with the following in mind: a party constructs a state $\rho_{ABC}$ by acting with the gates specified in a given circuit, and wants to know whether it is possible, if system $A$ is lost, to recover the state when given access to system $C$ only.  A YES-instance of this problem then corresponds to a recoverable state, since by the definition of fidelity of recovery in  \eqref{eq:fidrec}, there exists a recovery channel $\R_{C\rightarrow AC}$ which acts on $\rho_{BC}$ and satisfies the recovery criteria.  A NO-instance implies that no such recovery channel exists. We note that this problem is distinct from deciding whether $\rho_{ABC}$ is recoverable starting from the specification of the density matrix, a problem which has been shown to be decidable in classical polynomial time via a semi-definite program \cite{BT15}.

In order to have a robust operational meaning, it is important for this decision problem to have an efficient verification strategy, so that another party is unable to convince the verifier that a state is recoverable, if in fact it is not.  The complexity class \cc{QIP}($k$), introduced in \cite{QIPAmp00,QIP03}, captures this concept.  A problem is said to be in \cc{QIP}($k$) if, given $k$ distinct quantum messages exchanged between a verifier and a computationally unbounded prover, the verifier will accept YES-instances and reject NO-instances with very high probability.  The prover will always try to make the verifier accept, regardless of whether the state in question is a YES- or NO-instance. To prove \cc{FoR} $\subseteq$ \cc{QIP}, we will show that $F(A;B|C)_{\rho}$ is equal to the maximum acceptance probability of the verifier in a particular quantum interactive proof system. If this is true, then we can immediately conclude that the probability of accepting a YES-instance is no smaller than $\alpha$, and the probability of accepting a NO-instance is no larger than $\beta$, satisfying the properties of a QIP system. These probabilities can then be amplified to be exponentially close to the extremes of one and zero, respectively, by employing parallel repetition for \cc{QIP} \cite{QIPAmp00}.

\begin{figure*}
\includegraphics[width=1\linewidth]{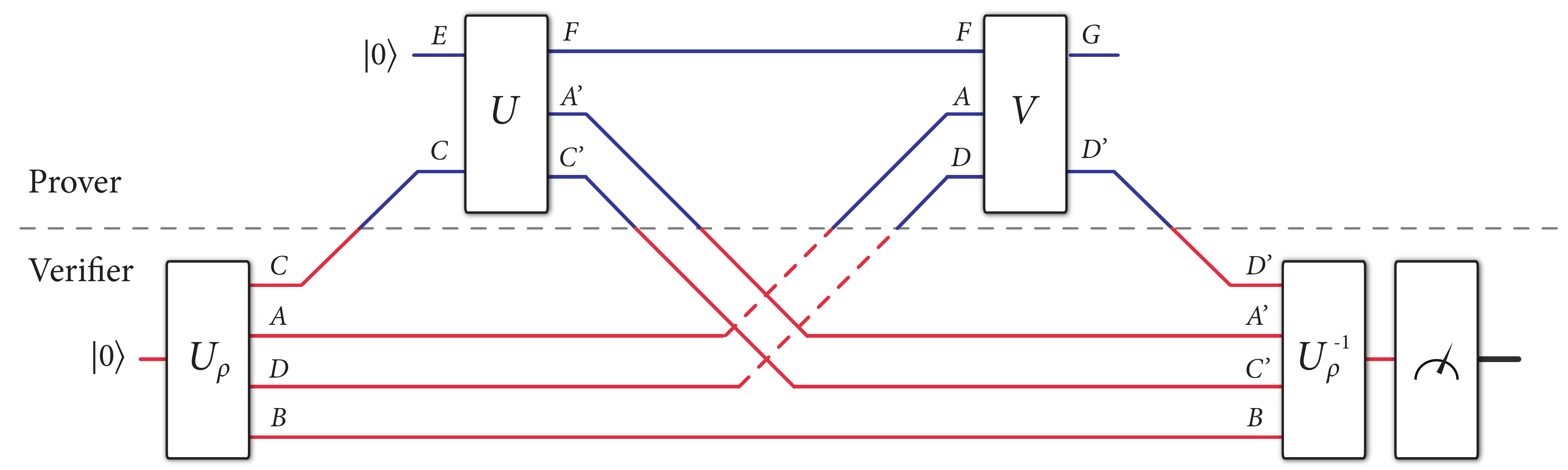}
\caption{This figure illustrates the quantum interactive proof system that establishes 1) an operational meaning of the fidelity of recovery and 2) the containment \cc{FoR} $\subseteq$ \cc{QIP}.  There are four distinct quantum messages exchanged between the verifier and the prover.} \label{fig:for}
\end{figure*}

We now give an outline of a quantum interactive proof system with maximum acceptance probability equal to the fidelity of recovery, thus witnessing the containment \cc{FoR} $\subseteq$ \cc{QIP}.  Recall that, for any pure four-party state $\phi_{ABCD}$, the fidelity of recovery obeys the following duality relation \cite{SW14}:  \begin{equation}
F(A;B|C)_{\phi}=F(A;B|D)_{\phi} . \label{eq:duality-FoR}
\end{equation}
The main idea behind this duality is that there is an optimal recovery channel for recovering $A$ from $C$ and an optimal dual ``Uhlmann'' recovery channel for recovering $A$ from $D$, and their performance as measured by fidelity is equal, as guaranteed by
Uhlmann's theorem \cite{U73}.
The proof system we construct is related to the methods used in \cite{SW14} to establish the relation
in \eqref{eq:duality-FoR}, but here we will have a computationally unbounded prover sequentially implement the recovery channel and the ``Uhlmann'' dual recovery channel \cite{U73}. In the setup of quantum interactive proofs, it is apparently necessary for such a prover to implement these channels given that the dimension of the Hilbert space is essentially exponentially large in $n$ (size of the circuit needed to generate $\rho_{ABC}$).  More explicitly, consider the following interaction between a verifier  and a prover, depicted in Figure~\ref{fig:for}:
\begin{enumerate}

\item The verifier uses the description of the quantum circuit to prepare the mixed state $\rho_{ABC}$ (with system $D$ a purifying system).   
In Figure~\ref{fig:for}, this is denoted by a unitary $U_{\rho}$ acting on many qubits prepared in the state $\vert 0 \rangle$, which we abbreviate simply as $\vert 0 \rangle$. The unitary $U_{\rho}$ has  output systems $A$, $B$, $C$, and $D$. So then
$\rho_{ABC} = \operatorname{Tr}_D\{  \vert \phi \rangle \langle \phi \vert_{ABCD} \} $ where $\vert \phi \rangle _{ABCD} \equiv
U_{\rho} \vert 0 \rangle$.

\item The verifier sends system $C$ to the prover.

\item The prover acts with a general unitary $U_{CE\rightarrow A'C'F}$ on system $C$ and an ancilla system $E$ prepared in a fiducial state $\vert 0 \rangle_E$, and the output systems are $F$, $A'$, and $C'$. 

\item The prover sends systems $A'$ and $C'$ to the verifier.  

\item The verifier sends systems $A$ and $D$ to the prover. 

\item The prover acts with a unitary $V_{ADF\rightarrow D'G}$ on systems $F$, $A$, and $D$ that has output  systems $D'$ and $G$. 

\item The prover sends $D'$ back to the verifier.

\item The verifier accepts if and only if the projection of the final state onto the pure state $\phi_{A'BC'D'}$ is successful. This test can be conducted by applying the inverse of the preparation unitary $U_{\rho}$, measuring each of the output qubits in the computational basis, and accepting if and only if the measurement outcomes are all zeros. 

\end{enumerate}

From this interaction, we can show via a chain of equalities that the maximum acceptance probability of the proof system is equal to the fidelity of recovery of the state $\rho_{ABC}$.
Consider that the maximum acceptance probability is equal to the following
Euclidean norm:%
\begin{align}
& \max_{U,V}\left\Vert \langle\phi|_{A^{\prime}BC^{\prime}D^{\prime}}V_{ADF\rightarrow
D^{\prime}G}U_{CE\rightarrow A^{\prime}C^{\prime}F}|\phi\rangle_{ABCD}%
|0\rangle_{E}\right\Vert _{2}^{2}\nonumber\\
& =\max_{U,V,|\varphi\rangle_{G}}\left\vert \langle\phi|_{A^{\prime}%
BC^{\prime}D^{\prime}}\langle\varphi|_{G}V U|\phi\rangle_{ABCD}|0\rangle_{E}\right\vert ^{2}\\
& =\max_{U,V}\left\vert \langle\phi|_{A^{\prime}BC^{\prime}D^{\prime}}\langle
\varphi|_{G}V U |\phi\rangle_{ABCD}|0\rangle_{E}\right\vert ^{2}, \label{eq:max-accept-prob-1}
\end{align}
where the first equality follows because there exists a unit vector $|\varphi
\rangle_{G}$ which achieves the norm and the second because the optimization
over $|\varphi\rangle_{G}$ can be absorbed into the optimization over the unitary $V_{ADF\rightarrow
D^{\prime}G}$. Consider that systems $F$, $A$, and $D$ purify the following state%
\begin{multline}
\operatorname{Tr}_{FAD}\{U\left(  |\phi\rangle\langle\phi|_{ABCD}%
\otimes|0\rangle\langle0|_{E}\right)  U^{\dag}\} \\ =\operatorname{Tr}%
_{F}\{U\left(  \rho_{BC}\otimes|0\rangle\langle0|_{E}\right)  U^{\dag}\},
\end{multline}
and systems $D^{\prime}$ and $G$ purify the following state:%
\begin{equation}
\operatorname{Tr}_{D^{\prime}G}\{|\phi\rangle\langle\phi|_{A^{\prime}BC^{\prime}%
D^{\prime}}\otimes|\varphi\rangle\langle\varphi|_{G}\}=\rho_{A^{\prime}BC^{\prime}}.
\end{equation}
Thus, by Uhlmann's theorem \cite{U73} with $V_{ADF\rightarrow D^{\prime}G}$ as the
Uhlmann unitary, it follows that \eqref{eq:max-accept-prob-1} is equal to%
\begin{multline}
\max_{U}F(\rho_{A^{\prime}BC^{\prime}},\operatorname{Tr}_{F}\{U\left(
\rho_{BC}\otimes|0\rangle\langle0|_{E}\right)  U^{\dag}\})\\
=\max_{\mathcal{R}_{C\rightarrow A^{\prime}C^{\prime}}}F(\rho_{A^{\prime
}BC^{\prime}},\mathcal{R}_{C\rightarrow A^{\prime}C^{\prime}}(\rho_{BC}))\\
=F(A;B|C)_\rho,
\end{multline}
where the first equality follows by the well known theorem of Stinespring \cite{S55}, which
states that any quantum channel can be realized by adjoining an ancilla
system, acting with a unitary, and tracing out a system.   This establishes an operational interpretation of fidelity of recovery as the maximum acceptance probability of our quantum interactive proof system for \cc{FoR}. By the reasoning given above, it follows that \cc{FoR} $\subseteq$ \cc{QIP}.

To establish that \cc{FoR} is hard for \cc{QSZK}, we need only consider that a special case of \cc{FoR} occurs when the $C$ system is trivial, in which case the recovery channel reduces to a preparation of a state on system $A$ and we then need to decide whether $\max_{\sigma_A} F(\rho_{AB}, \sigma_A \otimes \rho_B)$ is above or below a given threshold. This problem however has already been shown in \cite{v011a003} to be \cc{QSZK}-complete, from which we  conclude that \cc{FoR} is hard for \cc{QSZK}.

\section{A two-message quantum interactive proof system for fidelity of recovery}

\label{sec:FoR-in-QIP2}

The quantum interactive proof system in Figure~\ref{fig:for} gives a direct operational interpretation of the fidelity of recovery in terms of its maximum acceptance probability.
However, from the perspective of computational complexity theory, the QIP system has more messages exchanged than are necessary. Indeed, a general result states that any QIP system can be parallelized to an equivalent one that has only three messages exchanged between the verifier and the prover \cite{QIPAmp00}. 

\begin{figure*}
\includegraphics[width=1\linewidth]{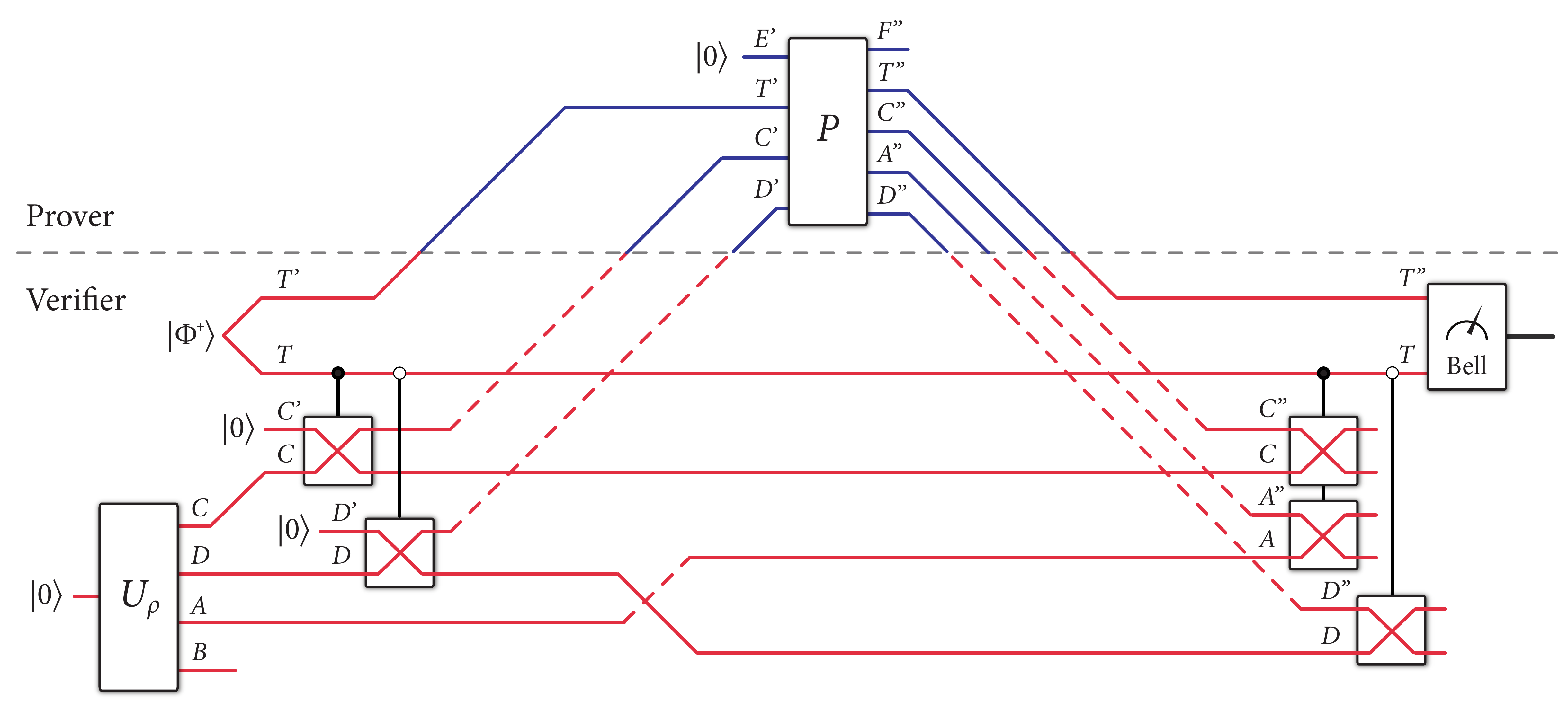}
\caption{A two-message quantum interactive proof system for deciding the fidelity of recovery computational problem.
The quantum gates with crossed wires denote controlled SWAP gates, as described in the main text. A ``filled-in'' circle indicates that the SWAP occurs controlled on the value in $T$ being equal to one, while a ``hollowed-out'' circle indicates that the SWAP occurs controlled on the value in $T$ being equal to zero.
} \label{fig:for-2}
\end{figure*}

In this section, we reduce the number of messages exchanged by showing that there exists a two-message quantum interactive proof system
for the fidelity of recovery computational problem.
By glancing at Figure~\ref{fig:for}, we see that the previous QIP system has the prover perform two actions: the recovery channel and the dual recovery channel, as discussed after \eqref{eq:duality-FoR}.
The idea of the two-message QIP system given in Figure~\ref{fig:for-2} is to force the prover to perform both actions in superposition. In terms of the many worlds interpretation of quantum mechanics, we can think that the verifier employs quantum entanglement and the superposition principle to force the prover to perform the recovery channel on system $C$ in one world, while in the other world redirecting the $D$ system to the prover so that he can perform the dual recovery channel on it. The verifier can then check at the end whether the prover took the correct actions in each world by realigning systems in each world, performing a Bell measurement, and demanding that the original entangled state prepared be undisturbed by the prover's actions. The result is that if the fidelity of recovery is high (so that by \eqref{eq:duality-FoR} both $F(A;B|C)$ and $F(A;B|D)$ are near to one), then there is a high probability that the verifier will be convinced that this is the case. If the fidelity of recovery is low, then there is little chance for the prover to convince the verifier.

We now detail the two-message QIP system.
Let $|\phi\rangle_{ABCD}$
denote\ the four-party pure state of interest. The proof system has the
following steps:

\begin{enumerate}
\item The verifier prepares a Bell state
\begin{equation}
|\Phi^{+}\rangle_{TT^{\prime}} 
\equiv \frac{1}{\sqrt{2}} (|00\rangle_{TT^{\prime}}+
|11\rangle_{TT^{\prime}}),
\end{equation}
the
four-party pure state $|\phi\rangle_{ABCD}$, and the ancilla states
$|0\rangle_{C^{\prime}}$ and $|0\rangle_{D^{\prime}}$.

\item The verifier performs a SWAP\ of $D$ and $D^{\prime}$ controlled on the
value in $T$ being equal to zero and a SWAP\ of $C$ and $C^{\prime}$
controlled on the value in $T$ being equal to one.

\item The verifier sends systems $T^{\prime}$, $C^{\prime}$, and $D^{\prime}$
to the prover.

\item The prover performs a quantum channel with systems $T^{\prime}$,
$C^{\prime}$, and $D^{\prime}$ as input and systems $T^{\prime\prime}$,
$A^{\prime\prime}$, $C^{\prime\prime}$, and $D^{\prime\prime}$ as output,
sending these back to the verifier. The output systems have the same size as
the corresponding input systems and system $A^{\prime\prime}$ has the same
size as system$\ A$. This quantum channel can be realized by adjoining an
ancilla $|0\rangle_{E^{\prime}}$ of sufficiently large size, performing a
unitary $P_{T^{\prime}C^{\prime}D^{\prime}E^{\prime}\rightarrow T^{\prime
\prime}C^{\prime\prime}D^{\prime\prime}A^{\prime\prime}F^{\prime\prime}}$ and
a partial trace over system $F^{\prime\prime}$.

\item The verifier performs a SWAP\ of $D$ and $D^{\prime\prime}$ controlled
on the value in $T$ being equal to zero and a SWAP\ of $C$ and $C^{\prime
\prime}$ controlled on the value in $T$ being equal to one. He also performs a
SWAP\ of $A$ and $A^{\prime\prime}$ controlled on the value in $T$ being equal
to one.

\item The verifier performs an incomplete Bell measurement on systems $T$ and
$T^{\prime\prime}$, with measurement operators
$\{ \Phi^{+}_{TT^{\prime\prime}}, I_{TT^{\prime\prime}} - \Phi^{+}_{TT^{\prime\prime}}\}$, and accepts if and only if the outcome is $\Phi^{+}_{TT^{\prime\prime}}$.
\end{enumerate}
Figure~\ref{fig:for-2} depicts this two-message QIP system.

We now analyze the maximum acceptance probability of this QIP system and show that it can never exceed a quantity related to the
fidelity of recovery. The acceptance probability given that the prover applies
a particular unitary $P_{T^{\prime}C^{\prime}D^{\prime}E^{\prime}\rightarrow
T^{\prime\prime}C^{\prime\prime}D^{\prime\prime}A^{\prime\prime}%
F^{\prime\prime}}$ is as follows:
\begin{widetext}
\begin{equation}
\left\Vert \langle\Phi^{+}|_{TT^{\prime\prime}}F_{T=1,AA^{\prime\prime}%
}F_{T=0,DD^{\prime\prime}}F_{T=1,CC^{\prime\prime}}P\ F_{T=0,DD^{\prime}%
}F_{T=1,CC^{\prime}}|\Phi^{+}\rangle_{TT^{\prime}}|\phi\rangle_{ABCD}%
|0\rangle_{C^{\prime}}|0\rangle_{D^{\prime}}|0\rangle_{E}\right\Vert _{2}^{2}%
\end{equation}
where we have abbreviated $P\equiv P_{T^{\prime}C^{\prime}D^{\prime}E^{\prime
}\rightarrow T^{\prime\prime}C^{\prime\prime}D^{\prime\prime}A^{\prime\prime
}F^{\prime\prime}}$ and $F_{T=1,CC^{\prime}}$ denotes a SWAP\ $C$ and
$C^{\prime}$ controlled on the value in $T$ being equal to one (with a similar
convention for the other controlled SWAP\ gates). We can then simplify the ket
in the above expression as%
\begin{align}
&  F_{T=1,AA^{\prime\prime}}F_{T=0,DD^{\prime\prime}}F_{T=1,CC^{\prime\prime}%
}P\ F_{T=0,DD^{\prime}}F_{T=1,CC^{\prime}}|\Phi^{+}\rangle_{TT^{\prime}}%
|\phi\rangle_{ABCD}|0\rangle_{C^{\prime}}|0\rangle_{D^{\prime}}|0\rangle
_{E}\nonumber\\
&  \propto F_{DD^{\prime\prime}}P_{T^{\prime}C^{\prime}D^{\prime}E^{\prime
}\rightarrow T^{\prime\prime}C^{\prime\prime}D^{\prime\prime}A^{\prime\prime
}F^{\prime\prime}}|0\rangle_{T}|0\rangle_{T^{\prime}}F_{DD^{\prime}}%
|\phi\rangle_{ABCD}|0\rangle_{C^{\prime}}|0\rangle_{D^{\prime}}|0\rangle
_{E^{\prime}}\nonumber\\
&  \qquad+F_{AA^{\prime\prime}}F_{CC^{\prime\prime}}P_{T^{\prime}C^{\prime
}D^{\prime}E^{\prime}\rightarrow T^{\prime\prime}C^{\prime\prime}%
D^{\prime\prime}A^{\prime\prime}F^{\prime\prime}}|1\rangle_{T}|1\rangle
_{T^{\prime}}F_{CC^{\prime}}|\phi\rangle_{ABCD}|0\rangle_{C^{\prime}}%
|0\rangle_{D^{\prime}}|0\rangle_{E^{\prime}}\\
&  =P_{T^{\prime}C^{\prime}D^{\prime}E^{\prime}\rightarrow T^{\prime\prime
}C^{\prime\prime}DA^{\prime\prime}F^{\prime\prime}}|0\rangle_{T}%
|0\rangle_{T^{\prime}}|\phi\rangle_{ABCD^{\prime}}|0\rangle_{C^{\prime}%
}|0\rangle_{D^{\prime\prime}}|0\rangle_{E^{\prime}}\nonumber\\
&  \qquad+P_{T^{\prime}C^{\prime}D^{\prime}E^{\prime}\rightarrow
T^{\prime\prime}CD^{\prime\prime}AF^{\prime\prime}}|1\rangle_{T}%
|1\rangle_{T^{\prime}}|\phi\rangle_{A^{\prime\prime}BC^{\prime}D}%
|0\rangle_{C^{\prime\prime}}|0\rangle_{D^{\prime}}|0\rangle_{E^{\prime}},
\end{align}
where $F_{DD^{\prime\prime}}$ denotes a SWAP of $D$ and $D^{\prime\prime}$
(with a similar convention for the other SWAP\ gates). Then the acceptance
probability simplifies as follows:%
\begin{equation}
\frac{1}{4}\left\Vert
\begin{array}
[c]{c}%
\langle0|_{T^{\prime\prime}}P_{T^{\prime}C^{\prime}D^{\prime}E^{\prime
}\rightarrow T^{\prime\prime}C^{\prime\prime}DA^{\prime\prime}F^{\prime\prime
}}|0\rangle_{T^{\prime}}|\phi\rangle_{ABCD^{\prime}}|0\rangle_{C^{\prime}%
}|0\rangle_{D^{\prime\prime}}|0\rangle_{E^{\prime}}\ \ 
+\\
\qquad\langle1|_{T^{\prime\prime}}P_{T^{\prime}C^{\prime}D^{\prime}E^{\prime
}\rightarrow T^{\prime\prime}CD^{\prime\prime}AF^{\prime\prime}}%
|1\rangle_{T^{\prime}}|\phi\rangle_{A^{\prime\prime}BC^{\prime}D}%
|0\rangle_{C^{\prime\prime}}|0\rangle_{D^{\prime}}|0\rangle_{E^{\prime}}%
\end{array}
\right\Vert _{2}^{2}.\label{eq:acc-prob-QIP-2}%
\end{equation}
The following two operators are contractions because $P_{T^{\prime}C^{\prime
}D^{\prime}E^{\prime}\rightarrow T^{\prime\prime}C^{\prime\prime}%
D^{\prime\prime}A^{\prime\prime}F^{\prime\prime}}$ is a unitary:%
\begin{align}
P_{C^{\prime}D^{\prime}E^{\prime}\rightarrow C^{\prime\prime}DA^{\prime\prime
}F^{\prime\prime}}^{00} &  \equiv\langle0|_{T^{\prime\prime}}P_{T^{\prime
}C^{\prime}D^{\prime}E^{\prime}\rightarrow T^{\prime\prime}C^{\prime\prime
}DA^{\prime\prime}F^{\prime\prime}}|0\rangle_{T^{\prime}},\\
P_{C^{\prime}D^{\prime}E^{\prime}\rightarrow CD^{\prime\prime}AF^{\prime
\prime}}^{11} &  \equiv\langle1|_{T^{\prime\prime}}P_{T^{\prime}C^{\prime
}D^{\prime}E^{\prime}\rightarrow T^{\prime\prime}CD^{\prime\prime}%
AF^{\prime\prime}}|1\rangle_{T^{\prime}}.
\end{align}
Then \eqref{eq:acc-prob-QIP-2} is equal to%
\begin{equation}
\frac{1}{4}\left\Vert P_{C^{\prime}D^{\prime}E^{\prime}\rightarrow
C^{\prime\prime}DA^{\prime\prime}F^{\prime\prime}}^{00}|\phi\rangle
_{ABCD^{\prime}}|0\rangle_{C^{\prime}}|0\rangle_{D^{\prime\prime}}%
|0\rangle_{E^{\prime}}+P_{C^{\prime}D^{\prime}E^{\prime}\rightarrow
CD^{\prime\prime}AF^{\prime\prime}}^{11}|\phi\rangle_{A^{\prime\prime
}BC^{\prime}D}|0\rangle_{C^{\prime\prime}}|0\rangle_{D^{\prime}}%
|0\rangle_{E^{\prime}}\right\Vert _{2}^{2}.\label{eq:acc-prob-QIP2-next}%
\end{equation}
Now consider for any two vectors $|\varphi_{1}\rangle$ and $|\varphi
_{2}\rangle$ that
$
\left\Vert |\varphi_{1}\rangle+|\varphi_{2}\rangle\right\Vert _{2}%
^{2}=\left\langle \varphi_{1}|\varphi_{1}\right\rangle +\left\langle
\varphi_{2}|\varphi_{2}\right\rangle +2\operatorname{Re}\{\left\langle
\varphi_{1}|\varphi_{2}\right\rangle \},
$
which implies that \eqref{eq:acc-prob-QIP2-next} is never larger than%
\begin{align}
&  \frac{1}{2}\left(  1+\operatorname{Re}\left\{  \langle\phi|_{ABCD^{\prime}%
}\langle0|_{C^{\prime}}\langle0|_{D^{\prime\prime}}\langle0|_{E^{\prime}%
}\left(  P_{C^{\prime}D^{\prime}E^{\prime}\rightarrow C^{\prime\prime
}DA^{\prime\prime}F^{\prime\prime}}^{00}\right)  ^{\dag}P_{C^{\prime}%
D^{\prime}E^{\prime}\rightarrow CD^{\prime\prime}AF^{\prime\prime}}^{11}%
|\phi\rangle_{A^{\prime\prime}BC^{\prime}D}|0\rangle_{C^{\prime\prime}%
}|0\rangle_{D^{\prime}}|0\rangle_{E^{\prime}}\right\}  \right)  \nonumber\\
&  =\frac{1}{2}+\frac{1}{2}\operatorname{Re}\left\{  \langle\phi|_{ABCD^{\prime}}\left[
\langle0|_{C^{\prime}}\langle0|_{E^{\prime}}\left(  P_{C^{\prime}D^{\prime
}E^{\prime}\rightarrow C^{\prime\prime}DA^{\prime\prime}F^{\prime\prime}}%
^{00}\right)  ^{\dag}|0\rangle_{C^{\prime\prime}}\right]  \left[
\langle0|_{D^{\prime\prime}}P_{C^{\prime}D^{\prime}E^{\prime}\rightarrow
CD^{\prime\prime}AF^{\prime\prime}}^{11}|0\rangle_{D^{\prime}}|0\rangle
_{E^{\prime}}\right]  |\phi\rangle_{A^{\prime\prime}BC^{\prime}D}\right\}  \nonumber\\
&  =\frac{1}{2}\left(  1+\operatorname{Re}\left\{  \langle\phi|_{ABCD^{\prime
}}\left(  V_{D^{\prime}\rightarrow A^{\prime\prime}DF^{\prime\prime}}\right)
^{\dag}U_{C^{\prime}\rightarrow CAF^{\prime\prime}}|\phi\rangle_{A^{\prime
\prime}BC^{\prime}D}\right\}  \right)  \nonumber\\
&  \leq\frac{1}{2}\left(  1+\left\vert \langle\phi|_{ABCD^{\prime}}\left(
V_{D^{\prime}\rightarrow A^{\prime\prime}DF^{\prime\prime}}\right)  ^{\dag
}U_{C^{\prime}\rightarrow CAF^{\prime\prime}}|\phi\rangle_{A^{\prime\prime
}BC^{\prime}D}\right\vert \right)  ,
\end{align}
where in the above we have defined the contractions%
\begin{equation}
V_{D^{\prime}\rightarrow A^{\prime\prime}DF^{\prime\prime}}   \equiv
\langle0|_{C^{\prime\prime}}P_{C^{\prime}D^{\prime}E^{\prime}\rightarrow
C^{\prime\prime}DA^{\prime\prime}F^{\prime\prime}}^{00}|0\rangle_{C^{\prime}%
}|0\rangle_{E^{\prime}},\qquad
U_{C^{\prime}\rightarrow CAF^{\prime\prime}}   \equiv\langle0|_{D^{\prime
\prime}}P_{C^{\prime}D^{\prime}E^{\prime}\rightarrow CD^{\prime\prime
}AF^{\prime\prime}}^{11}|0\rangle_{D^{\prime}}|0\rangle_{E^{\prime}}.
\end{equation}
Consider that%
\begin{multline}
  \left\vert \langle\phi|_{ABCD^{\prime}}\left(  V_{D^{\prime}\rightarrow
A^{\prime\prime}DF^{\prime\prime}}\right)  ^{\dag}U_{C^{\prime}\rightarrow
CAF^{\prime\prime}}|\phi\rangle_{A^{\prime\prime}BC^{\prime}D}\right\vert
\\
  \leq\max_{V,U}\left\{  \left\vert \langle\phi|_{ABCD^{\prime}}\left(
V_{D^{\prime}\rightarrow A^{\prime\prime}DF^{\prime\prime}}\right)  ^{\dag
}U_{C^{\prime}\rightarrow CAF^{\prime\prime}}|\phi\rangle_{A^{\prime\prime
}BC^{\prime}D}\right\vert :\left\Vert V\right\Vert _{\infty},\left\Vert
U\right\Vert _{\infty}\leq1\right\}    =\sqrt{F}(A;B|C)_{\phi}.
\end{multline}
\end{widetext}
where the last equality follows from the duality of fidelity of recovery and
because any contraction can be written as a convex combination of isometries,
so that there is an optimal pair of isometries achieving the maximum in the
second line \cite[Theorem~5.10]{Z11}. Thus, the maximum acceptance probability for the
QIP system is never higher than%
\begin{equation}
\frac{1}{2}\left(  1+\sqrt{F}(A;B|C)_{\phi}\right)  .
\label{eq:QIP2-accept-prob}
\end{equation}
This upper bound on the acceptance probability can be achieved if the prover
applies a unitary extension of the following isometry:%
\begin{multline}
P_{T^{\prime}C^{\prime}D^{\prime}E^{\prime}\rightarrow T^{\prime\prime
}CD^{\prime\prime}AF^{\prime\prime}}=
\\|0\rangle_{T^{\prime\prime}}%
\langle0|_{T^{\prime}}\otimes V_{D^{\prime}\rightarrow A^{\prime\prime
}DF^{\prime\prime}}|0\rangle_{C^{\prime\prime}}\langle0|_{C^{\prime}}%
\langle0|_{E^{\prime}}\\
+|1\rangle_{T^{\prime\prime}}\langle1|_{T^{\prime}}\otimes U_{C^{\prime
}\rightarrow CAF^{\prime\prime}}|0\rangle_{D^{\prime\prime}}\langle
0|_{D^{\prime}}\langle0|_{E^{\prime}},
\end{multline}
where $V_{D^{\prime}\rightarrow A^{\prime\prime}DF^{\prime\prime}}$ and
$U_{C^{\prime}\rightarrow CAF^{\prime\prime}}$ are isometries achieving the
maximum in the fidelity of recovery $F(A;B|C)_{\phi}$.

Thus, in the case of a YES instance, there exists a strategy to convince the verifier to accept with the probability in \eqref{eq:QIP2-accept-prob}, while in the case of a NO instance, no strategy can convince the verifier to accept with probability higher than that in \eqref{eq:QIP2-accept-prob}. Given the promise from Problem~\ref{prob:FoR} and known error reduction procedures for \cc{QIP}(2) \cite{JUW09}, these probabilities can then be amplified to be exponentially close to the extremes of one and zero, respectively.

\section{Operational meaning of regularized relative entropy of recovery}

In this section, we provide an operational interpretation of the regularized relative entropy of recovery in the context of quantum hypothesis testing \cite{HP91,ON00}. The setting is as discussed in the introduction:\ Given are $n$ copies of a state $\rho_{ABC}$, and the task is to determine whether $\rho_{ABC}^{\otimes n}$ is prepared or whether $\mathcal{R}_{C^{n}\rightarrow A^{n}C^{n}}(  \rho_{BC}^{\otimes n})  $ is prepared, where $\mathcal{R}_{C^{n}\rightarrow A^{n}C^{n}}$ is some recovery channel. This is an instance of a more general problem of discriminating between a state $\rho^{\otimes n}$ and a set $\mathcal{S}^{\left(  n\right)  }$ of states, where in our case:
\begin{align}
\rho^{\otimes n}  &  =\rho_{ABC}^{\otimes n},\\
\mathcal{S}^{\left(  n\right)  }  &  =\left\{  \mathcal{R}_{C^{n}\rightarrow A^{n}C^{n}}(  \rho_{BC}^{\otimes n})  :\mathcal{R}\in\text{CPTP}\right\}  , \label{eq:set-S-recover}
\end{align}
with CPTP denoting the set of quantum channels from $C^n$ to $A^n C^n$.
This more general setting was studied in detail in \cite{BP10}, where it was found that the Type II\ rate of convergence simplifies if the following conditions hold:

\begin{enumerate}
\item (Convexity) $\mathcal{S}^{\left(  n\right)  }$ is convex and closed for all $n$.

\item (Full Rank) There exists a full rank state $\sigma$ such that each $\mathcal{S}^{\left(  n\right)  }$ contains $\sigma^{\otimes n}$.

\item (Reduction) For each $\sigma\in\mathcal{S}^{\left(  n\right)  }$, Tr$_{n}\{\sigma\}  \in\mathcal{S}^{\left(  n-1\right)  }$.

\item (Concatenation) If $\sigma_{n}\in\mathcal{S}^{\left(  n\right)  }$ and $\sigma_{m}\in\mathcal{S}^{\left(  m\right)  }$, then $\sigma_{n}\otimes\sigma_{m} \in\mathcal{S}^{\left(  n+m\right)  }$.

\item (Permutation invariance) $\mathcal{S}^{\left(  n\right)  }$ is closed under permutations.
\end{enumerate}

We now verify that the set $\mathcal{S}^{\left(  n\right)  }$ as defined in (\ref{eq:set-S-recover}) satisfies the above properties.

\textbf{Convexity}. Let $\mathcal{R}_{C^{n}\rightarrow A^{n}C^{n}}^{1}(  \rho _{BC}^{\otimes n})  ,\mathcal{R}_{C^{n}\rightarrow A^{n}C^{n}}%
^{2}(  \rho_{BC}^{\otimes n})  \in\mathcal{S}^{\left(  n\right)  } $. Then for all $\lambda\in\left[  0,1\right]  $, we have that%
\begin{equation*}
\lambda\mathcal{R}_{C^{n}\rightarrow A^{n}C^{n}}^{1}(  \rho_{BC}^{\otimes n})  +\left(  1-\lambda\right)  \mathcal{R}_{C^{n}\rightarrow A^{n}%
C^{n}}^{2}(  \rho_{BC}^{\otimes n})  \in\mathcal{S}^{\left( n\right)  }%
\end{equation*}
because $\lambda\mathcal{R}_{C^{n}\rightarrow A^{n}C^{n}}^{1}+\left(1-\lambda\right)  \mathcal{R}_{C^{n}\rightarrow A^{n}C^{n}}^{2}$ is a quantum channel if
$\mathcal{R}_{C^{n}\rightarrow A^{n}C^{n}}^{1}$ and $\mathcal{R}_{C^{n}\rightarrow A^{n}C^{n}}
^{2}$ are. Furthermore, the set of all CPTP\ maps is closed.

\textbf{Full Rank}. Without loss of generality, we can assume that $\rho_{B}$ is a full rank state. A particular recovery channel is one which traces out system $C$ and replaces with the maximally mixed state on $AC$. Taking $n$ copies of such a state gives a full-rank state in $\mathcal{S}^{\left(  n\right)  }$.

\textbf{Reduction}. Let $\mathcal{R}_{C^{n}\rightarrow A^{n}C^{n}}(  \rho_{BC}^{\otimes n})  \in\mathcal{S}^{\left(  n\right)  }$. Consider that%
\begin{multline}
\text{Tr}_{A_{n}B_{n}C_{n}}\{  \mathcal{R}_{C^{n}\rightarrow A^{n}C^{n}%
}(  \rho_{BC}^{\otimes n})  \}  = \\ \text{Tr}_{A_{n}C_{n}%
}\{  \mathcal{R}_{C^{n}\rightarrow A^{n}C^{n}}(  \rho_{BC}^{\otimes n-1}\otimes\rho_{C})  \}  .
\end{multline}
This state is in $\mathcal{S}^{\left(  n\right)  }$ because the recovery channel for $\rho_{BC}^{\otimes n-1}$ could consist of tensoring in $\rho_{C}$, applying $\mathcal{R}_{C^{n}\rightarrow A^{n}C^{n}}$, and tracing out systems $A_{n}C_{n}$.

\textbf{Concatenation}. Let $\mathcal{R}_{C^{n}\rightarrow A^{n}C^{n}}^{1}(  \rho _{BC}^{\otimes n})  \in\mathcal{S}^{\left(  n\right)  }$ and $\mathcal{R}_{C^{m}\rightarrow A^{m}C^{m}}^{2}(  \rho_{BC}^{\otimes m})  \in\mathcal{S}^{\left(  m\right)  }$. Then
\begin{equation}
\mathcal{R}%
_{C^{n}\rightarrow A^{n}C^{n}}^{1}(  \rho_{BC}^{\otimes n}) \otimes\mathcal{R}_{C^{m}\rightarrow A^{m}C^{m}}^{2}(  \rho_{BC}^{\otimes m})  \in\mathcal{S}^{\left(  n+m\right)  },
\end{equation}
 because
\begin{multline}
\mathcal{R}_{C^{n}\rightarrow A^{n}C^{n}}^{1}(  \rho_{BC}^{\otimes n})  \otimes\mathcal{R}_{C^{m}\rightarrow A^{m}C^{m}}^{2}(
\rho_{BC}^{\otimes m})  = \\ \left(  \mathcal{R}_{C^{n}\rightarrow A^{n}C^{n}}^{1}\otimes\mathcal{R}_{C^{m}\rightarrow A^{m}C^{m}}^{2}\right)
(  \rho_{BC}^{\otimes n+m})  ,
\end{multline}
so that the recovery channel consists of the parallel concatenation of
$\mathcal{R}_{C^{n}\rightarrow A^{n}C^{n}}^{1}$ and $\mathcal{R}%
_{C^{m}\rightarrow A^{m}C^{m}}^{2}$.

\textbf{Permutation invariance}. Here, we need to show that for $\sigma\in\mathcal{S}^{\left(  n\right) }$, we have that $\pi\sigma\pi^{\dag}\in\mathcal{S}^{\left(  n\right)  }$ for
all permutations $\pi$ of the $n$ systems. Let $\mathcal{R}_{C^{n}\rightarrow A^{n}C^{n}}(  \rho_{BC}^{\otimes n})  \in\mathcal{S}^{\left(
n\right)  }$. Then
\begin{align}
&  \pi_{A^{n}B^{n}C^{n}}\mathcal{R}_{C^{n}\rightarrow A^{n}C^{n}}( \rho_{BC}^{\otimes n})  \left(  \pi_{A^{n}B^{n}C^{n}}\right)  ^{\dag
}\nonumber\\
&  =\left(  \pi_{A^{n}}\otimes\pi_{B^{n}}\otimes\pi_{C^{n}}\right) \mathcal{R}
(  \rho_{BC}^{\otimes n})
\left(  \pi_{A^{n}}\otimes\pi_{B^{n}}\otimes\pi_{C^{n}}\right)  ^{\dag}\nonumber\\
&  =\left(  \pi_{A^{n}}\otimes\pi_{C^{n}}\right)  \mathcal{R}
(  \pi_{B^{n}}\rho_{BC}^{\otimes n}\pi_{B^{n}%
}^{\dag})  \left(  \pi_{A^{n}}\otimes\pi_{C^{n}}\right)  ^{\dag}\nonumber\\
&  =\left(  \pi_{A^{n}}\otimes\pi_{C^{n}}\right)  \left[  \mathcal{R}%
(  \pi_{C^{n}}^{\dag}\rho_{BC}^{\otimes n}\pi_{C^{n}})  \right]  \left(  \pi_{A^{n}}\otimes\pi_{C^{n}}\right)
^{\dag}  \nonumber\\
& \in\mathcal{S}^{\left(  n\right)  },
\end{align}
where the second equality follows because the permutation of the $B$ systems commutes with the recovery channel, the third because $\rho_{BC}^{\otimes n}$ is a permutation invariant state, and the last line because a potential recovery consists of applying the permutation $\pi_{C^{n}}^{\dag}$, followed by $\mathcal{R}_{C^{n}%
\rightarrow A^{n}C^{n}}$, followed by the permutation $\pi_{A^{n}}\otimes \pi_{C^{n}}$.

From here, we can define a hypothesis testing relative entropy of recovery for a state $\rho_{ABC}$ as follows:%
\begin{equation*}
D_{H}^{\varepsilon}(  A;B|C)  _{\rho}\equiv\inf_{\mathcal{R}%
_{C\rightarrow AC}}D_{H}^{\varepsilon}(  \rho_{ABC}\Vert\mathcal{R}%
_{C\rightarrow AC}(  \rho_{BC})  )  ,
\end{equation*}
where $D_{H}^{\varepsilon}$ is the hypothesis testing relative entropy \cite{BD10,WR10}, defined for two states $\omega$ and $\tau$ as
\begin{multline*}
D_H^{\varepsilon}(\omega \Vert \tau) \equiv \\
-\log \min_{Q} \{ \operatorname{Tr} \{Q\tau\} :
0 \leq Q \leq I \wedge \operatorname{Tr}\{Q \omega\} \geq 1-\varepsilon \} .
\end{multline*}
By definition, the hypothesis testing relative entropy 
$D_{H}^{\varepsilon}$
is equal to the optimal Type II error exponent when the Type I error cannot exceed $\varepsilon \in (0,1)$. By employing the main result of \cite{BP10} and the above
observations, we can conclude that%
\begin{equation}
\lim_{n\rightarrow\infty}\frac{1}{n}D_{H}^{\varepsilon}(  A^{n}%
;B^{n}|C^{n})  _{\rho^{\otimes n}}=D^{\infty}(  A;B|C)_{\rho}, \label{eq:key-identity}%
\end{equation}
for all $\varepsilon\in\left(  0,1\right)  $.
(Note that the limit as $\varepsilon\rightarrow0$ is not needed.)\ This gives an operational interpretation of $D^{\infty}(  A;B|C)  _{\rho}$ as the optimal Type II error exponent as claimed.

\section{Conclusion} \label{sec:conclusion}

We have given operational meaning to two different recovery measures: the fidelity of recovery and the relative entropy of recovery. The first occurs in a ``one-shot'' scenario, where we find that the fidelity of recovery is equal to the maximum probability with which a quantum prover can convince a quantum verifier that a given state is recoverable. As an additional contribution, we give a different quantum interactive proof system for the fidelity of recovery problem which has only two messages exchanged between the verifier and the prover. Thus we make progress on a computational problem related to recoverability by showing that the problem \cc{FoR} is in \cc{QIP}(2) and is hard for \cc{QSZK}. The second operational interpretation occurs in a scenario involving many copies of a given tripartite state and represents a generalization of quantum Stein's lemma \cite{HP91,ON00}. We showed that the optimal Type II error exponent is equal to the regularized relative entropy of recovery if there is a constraint on the Type I error.

Going forward from here, it would be interesting to give better bounds on the computational problem \cc{FoR}. For example, could we show that \cc{FoR} is hard for \cc{QIP}(2)? For the hypothesis testing setup, can we give finer characterizations of the optimal Type II exponent when the Type I error is not a fixed constant but decays as well (cf., \cite{MO15})? Perhaps the R\'enyi relative entropy of recovery studied in \cite{BFT15} would be relevant here? This question was recently addressed and solved in the specialized classical case \cite{TH15}, but additivity issues pose a significant challenge to extending results like these to the quantum case.

\begin{acknowledgments}
We are grateful to Mario Berta, Omar Fawzi, and Marco Tomamichel for discussions and for sharing an early draft of \cite{BFT15}. We also thank Patrick Hayden for discussions. CH is supported by the Spanish MINECO, project FIS2013-40627-P, as well as by the Generalitat de Catalunya, CIRIT project no.~2014 SGR 966.
JW is supported in part by Canada's NSERC.
MMW acknowledges support from the NSF under Award No.~CCF-1350397.

\end{acknowledgments}

\bibliographystyle{unsrt}
\bibliography{Ref}

\end{document}